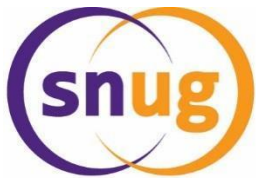

**Synopsys Users Group**

# Standard Cell Library Evaluation with Multiple-lithography-compliant verification and Improved Synopsys Pin Access Checking Utility

Yongfu Li, Wan Chia Ang, Chin Hui Lee,
Kok Peng Chua, Yoong Seang Jonathan Ong, Chiu Wing Colin Hui

GLOBALFOUNDRIES
Singapore

**ABSTRACT**

While standard cell layouts are drawn with minimum design rules to maximize the benefit of design area shrinkage, the complicated design rules have caused difficulties with signal routes accessing the pins in standard cell layouts. As a result, it has become a great challenge for physical layout designers to design a standard cell layout that is optimized for area, power, timing, signal integrity, and printability. Multiple design iterations are required to consider pin accessibility during standard cells layout to increase the number of feasible solutions available to the router. In this work, we will demonstrate several improvements with the Synopsys PAC methodology, such as reducing the number of cells required for each Synopsys 'testcell' with the same cell abutment condition, increasing the complexity of the pin connection for better pin accessibility evaluation. We also recommend additional constraints to improve the probability of detecting pin accessibility issues. We also integrate other physical verification methods to access the design rule compliance and the printability of standard cells. We hope that the easy to use utility enables layout engineers to perform the verification, simplifying the verification methodology.

# Table of Contents

# Table of Figures

# Table of Tables

*Standard Cell Library Evaluation with Multiple-lithography-compliant
verification and Improved Synopsys Pin Access Checking Utility*

# 1   Introduction

The continuous scaling of the CMOS technology is now facing the bottleneck of using 193i optical lithography process. Pattern-related defects continue to increase and limit the number of good die per wafer. The classical rule-based Design Rule Check (DRC) approach is no longer sufficient to guarantee 100% pattern printability. Starting from the GLOBALFOUNDRIES 40-nm technology, Design-for-Manufacturability (DFM) verifications are required to identify manufacturing weak-points and prevent catastrophic errors such as open (necking) and shorts (bridging) issues, thus enabling early ramp to good yield [1], [2].

With the additional of physical verifications in every new technology generation, the development resource for the intellectual property (IP) libraries has become astronomical challenging and costly. In particular, standard cell IP libraries are the first and foremost design foundation libraries to construct the place-and-route (P&R) digital circuit design; the quality of the standard cell layouts has a direct impact on the chip design's area and manufacturability. Today, we observe that the key challenges limiting the quality of the standard cell library are the placement of the pin locations and lithography-compliant (lithography printability and color-compliant) design, respectively. Therefore, an early assessment of the design restrictions imposed by the circuit design is required.

## 1.1   Challenge #1: Pin accessibility in standard cell

As the standard cell layouts are drawn with minimum design rules to maximize the benefit of design area shrinkage, the number of routing tracks has begun to decrease from 12-track in 65-nm to 7.5-track in 14-nm. As a result, the modern router tools have difficulties with signal routes accessing to the standard cell pins [3]. In particular, Hsu, et al. reported that the challenges with pin-access have severely degraded routing resource estimation accuracy with the convention global and detail routing model [4]. Multiple design iterations are required to resolve the routing congestion and pin accessibility issues, or even required in a change in the design core utilization, thus resulting in an increase in the overall chip area [5]. The experiment has shown that the smallest chip area may not be easily achieved by standard cells with the smallest area [3].

To achieve an optimal design and to mitigate the routing congestion, it is important to consider pin accessibility during the layout development phase. Synopsys has introduced the Synopsys Pin Access Checker (PAC) utility in the library preparation reference methodology, which aims to identify problematic pin location in the standard cell layouts [6]. The technique motivates us to explore and to improve the detection coverage with our proposed design constraints with the utility tool.

## 1.2   Challenge #2: lithography-compliant standard cell

A combination of DRC and lithography simulations is required to ensure a fully lithography-compliant (lithography printability and color-compliant) standard cells library. From our experience, we have observed that coloring and lithography violations do occur at the cell boundaries mainly because of the insufficient distance of the metal layers from the boundary. Although color spacing can be checked during the placement stage to prevent DP violations due to standard cell abutment [7], there is no existing integrated tool to address both coloring and lithography printability for the entire standard cell boundaries. With the number of standard cells increases with various design variants and functionality, the total number of cell placement combinations grows exponentially. More combinations yield more physical area and polygons which impact the verification run time.

Therefore, to address the critical challenges in standard cell IP qualification, we propose and demonstrate an integrated verification tool to achieve the following objectives:
  (a) To design a simplified abutment unit testcase to eliminates redundant cases and maintains 100% coverage of all standard cell abutment topologies with minimum area.
  (b) To improve the detection coverage with our proposed design constraints with the Synopsys PAC utility tool.
  (c) To assist layout engineer to identify lithography weak-points at the standard cell abutment boundaries so that they can address the errors.

The rest of this paper is organized as follows. Section II explains the present methodology in identifying the pin accessibility and lithography printability issues. Section III presents the overview of the Synopsys PAC flow. Section IV presents our proposed simplified abutment unit testcase. Section V presents several methods to increase the detection coverage for problematic pins in the standard cell libraries. Section VI details the implementation of our Tcl procedure script. Finally, Section VII presents the result using our proposed utility tool and conclusions are given in Section VIII.

## 2   Present standard cell library design methodology and the challenge with identifying pin accessibility & lithography printability issues

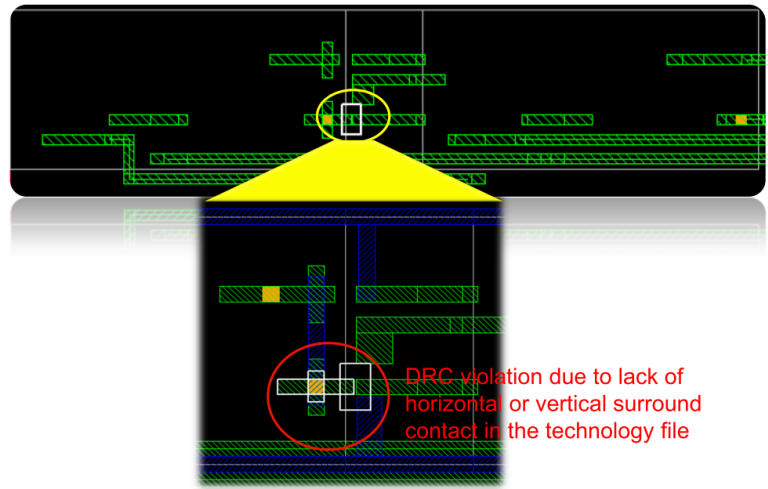

Figure 1. Metal 2 (M2) Spacing violation due to M2 – Via one (V1) inserted at the pin location.

The design and qualification of the standard cell library involve various engineering groups such as library development engineers, layout engineers, digital circuit designers and physical verification engineers.  The entire process can be summarized as follow:
(a) The library development team creates a variety of digital logic circuit function.
(b) The layout engineers implement the drawing under the constraints of minimum design rules, fixed cell height, and an integer multiple of routing tracks.
(c) The verification engineers perform various verifications to ensure the functionality and manufacturability of the library.

As shown in Figure 1,  to avoid spacing violation at the standard cell boundary due to the insertion of the metal two and via one at the pin location, the library development team has to perform physical verification checks on a P&R digital circuit design in order to identify and fix the standard cells with pin-access difficulties. This involves the characterization of the standard cells' layout, followed by the generation of the Synopsys timing liberty library format.  Additional resources and time are required to perform multiple iterations for any changes in the layout or any new cells added to the library. Another disadvantage is that the synthesized netlist might not achieve hundred percentage standard cells' coverage. Thus, the solution is less likely to be adopted for incremental development work.

There is a need to simplify the present solution and provides an easy-to-use utility for library physical layout designer to perform all the tasks with hundred percent verification coverage. The simple-to-use approach reduces the barriers to adopting new tool, especially the layout engineers who may not be well-versed with programming and digital design flow.

# 3    Synopsys Pin Access Checker (PAC) Flow

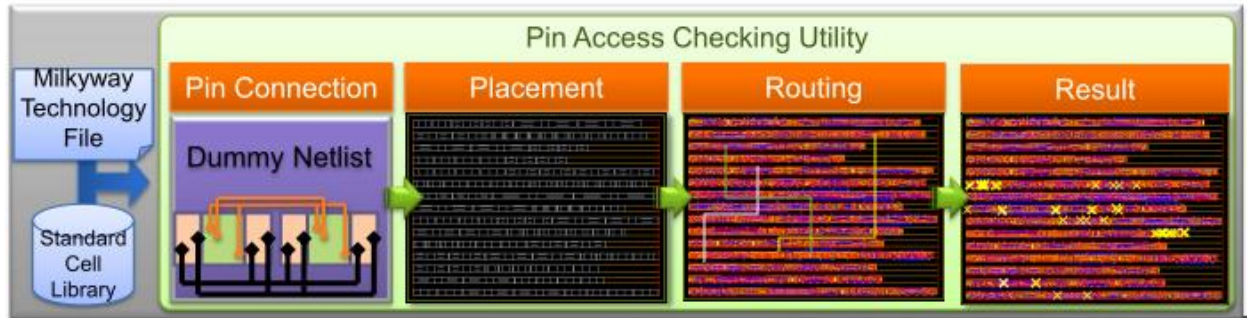

Figure 2. Synopsys PAC utility workflow.

To address the pin accessibility challenge with a simplified solution, Synopsys has provided the Synopsys PAC utility tool as part of the Synopsys IC Compiler (ICC) library preparation reference methodology [6], [7], and the concept of the utility tool is summarized in Figure 2. The utility enumerates all combinations of the standard cells and generates a Verilog gate-level netlist and a design exchange format (DEF) file which is essential to define the logical connections and physical placements of the generated design, respectively. Synopsys termed this circuit-like design as 'testcell' method. The utility also leverages on the Synopsys ICC zroute capability to perform automated routing and design rule check (DRC) verification. The DRC violations help to identify the pin locations associated with routing congestion issues. Therefore, this utility simplifies the physical verification flow for pin accessibility check without the need for timing library and timing constraints.

The Synopsys PAC utility provides four different modes of checking pin accessibility, namely (1) single_cell_only, (2) cell_by_cell_only, (3) all_combo_in_one_cell_only and (4) all [6], [7]. The first two modes allow library development engineers to perform incremental verification during the standard cell layout development phase and the latter two modes can be performed on the final library before silicon validation phase. The latter two modes create the top cell with the entire individual 'testcells' generated through the modes (1) and (2). The multiple-tier verification approach helps to perform verification with limited runtime overhead at the various phases of library development. Thus, the library development engineers will be able to correct the problematic cells before silicon validation phase.

In [7], Aupoix, et al. evaluated the Synopsys PAC utility with its default constraints on several standard cell libraries, which only enables them to identify one problematic pin issue among the libraries. Furthermore, the initial pin accessibility problem with the testcell was later resolved with rerouting. As the detail router is based on the heuristic algorithm, the final routed design highly depends on the placement of the standard cells and the number of routing iterations. Therefore, the number of problematic pins identified through this methodology depends on the final routed design.

Based on the previous learnings in [7], we attempt to explore several methods to constraints the placement of 'testcell' and it's routing, thus exacerbating the pin accessibility issue. With the proposed methods, we aim to improve the probability of detecting pin accessibility issues.

## 4 Unit Testcase using reduction techniques for standard cell abutment placement

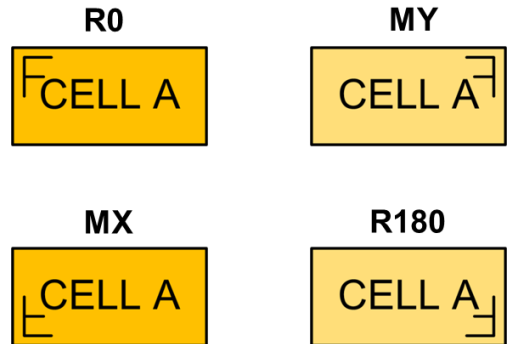

Figure 3. Four possible legal orientations of standard cell placement in the chip design

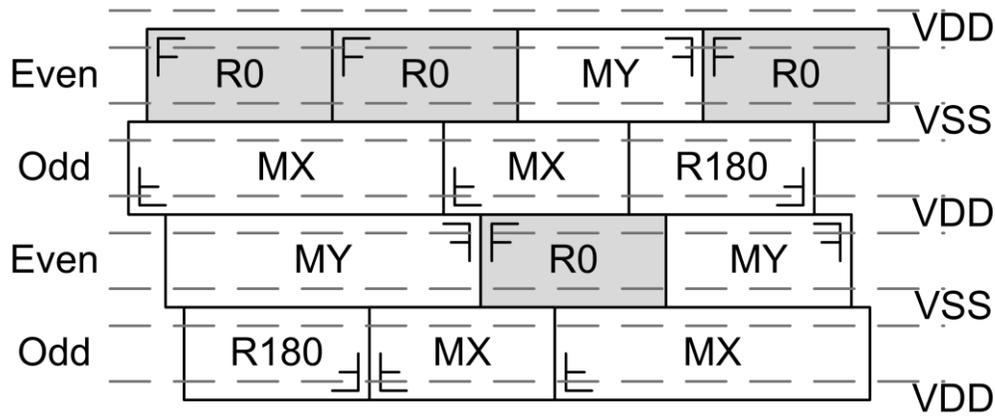

Figure 4. Standard cells placement in the chip design

In the chip design, there are four possible orientations of a standard cell that can be placed in the layout, as shown in Figure 3. The R0- and R180-orientated cells are the default cell orientation and cell rotated by 180 degree. The MX- and MY-orientated cells are cells mirrored along the X-axis and Y-axis, respectively. A conventional single-height standard cell is designed with the power signal (VDD) and the ground signal (VSS) at the top and bottom rails, respectively. Therefore, the single-height standard cells with R0- and MY-orientations are legally placed at the same row while the single-height standard cells with MX- and R180-orientations are placed at the alternate row. This arrangement allows the standard cells to be placed along pre-defined, interleaved horizontal VDD and VSS rails during placement phase. As shown in Figure 4, a standard cell (highlighted in 'grey') can be placed in the layout with orientation of R0 and be surrounded by other cells with orientation of R0 or MY.

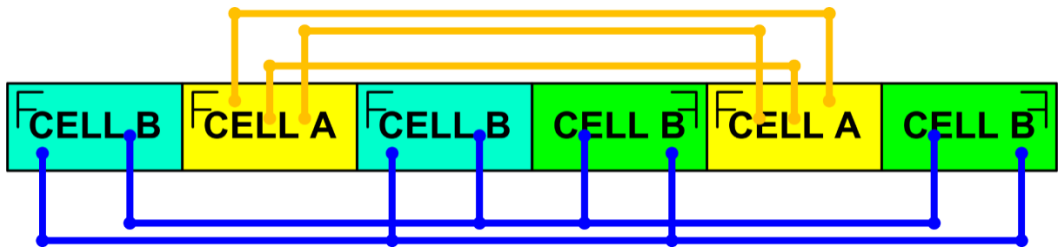

Figure 5. The schematic illustrates the simplified version of the Synopsys 'testcell' method

Synopsys has introduced the 'testcell' method, as illustrated in Figure 5, where the target cell A (with R0 orientation) is placed in the middle with two cell B (with R0 orientations) are abutted on its right and left sides and the second set is similar to the first set, but the two cell B are flipped in the horizontal direction (with MY orientations). Thus, the placement in the Synopsys 'testcell' has considered all the possible combinations using the minimum number of cell required.

Although the Synopsys 'testcell' provides 1.3 times reduction in the number of cells compared to the conventional method, there is a need to further optimize the testcell in order to reduce the database and runtime significantly. Without the loss of generality, we proposed three types of area-efficient cell abutment placements, which reduce the number of required standard cells compared to the Synopsys 'testcell'. In order to achieve a significant cell reduction, we have to consider the following scenarios:

- Single-height type A-A – Single-height side-to-side standard cell abutment with identical cells.
- Single-height type A-B – Single-height side-to-side standard cell abutment with different cells.
- Single-multiple-height type A-B – Single-height side-to-side standard cell abutment with the multiple-height cell.

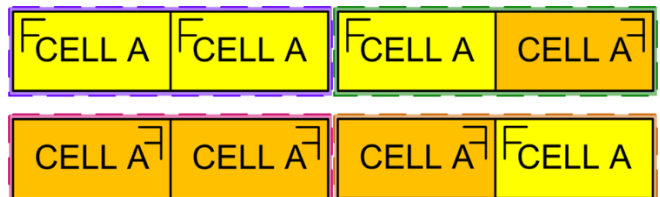

Figure 6. Four possible permutation pairs of placement for single-height type A-A condition

As shown in Figure 6, there are a total of four possible permutation pairs of placement for single-height type A-A condition. The number of standard cells can be reduced by overlapping and placing the entire permutated pairs in a single row.

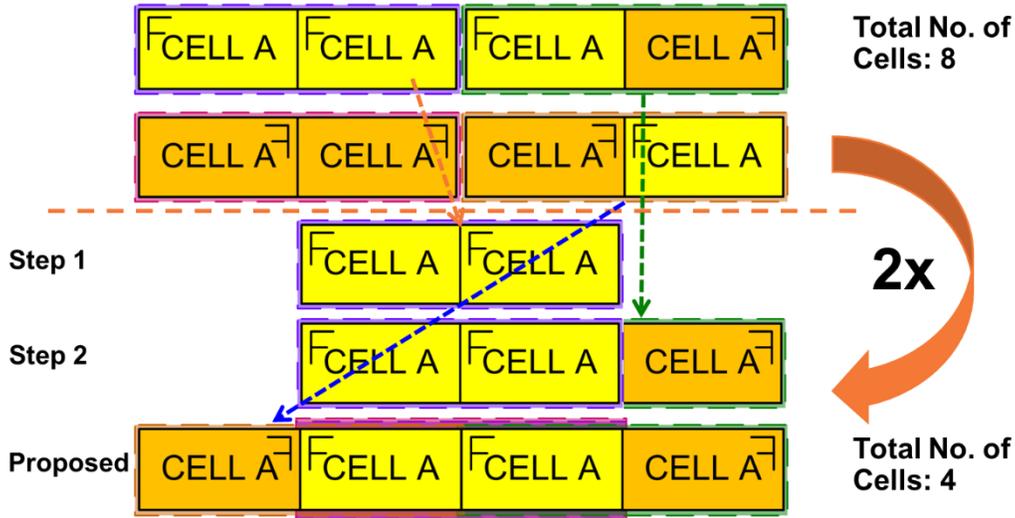

Figure 7. The proposed single-height type A-A cell abutment placement

As shown in Figure 7, the R0-R0 permutation pair overlaps with R0-MY and MY-R0 permutation pairs on the right and left sides, respectively. The MY-MY permutation pair is discarded in the proposed abutment placement because it is a mirror version of R0-R0. The proposed abutment placement can achieve an area reduction of 2 times.

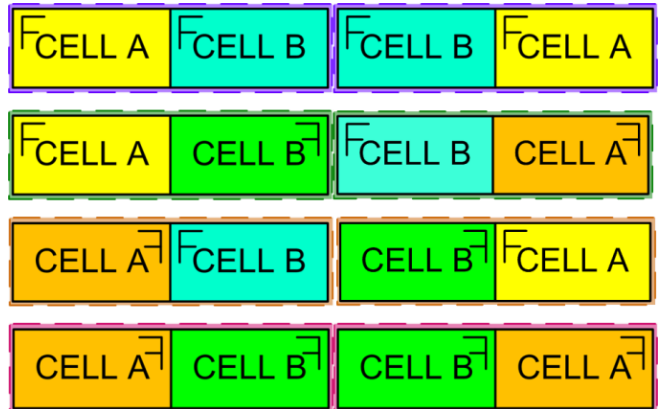

Figure 8. Eight possible permutation pairs of placement for single-height type A-B condition

For single-height type A-B condition, there are a total of eight possible permutation pairs of placement, as shown in Figure 8. As illustrated in Figure 9, using the similar technique described earlier, the number of standard cells can be reduced by overlapping and placing the entire permutated pairs in a single row. The proposed abutment placement can achieve an area reduction of 3.2 times. Furthermore, there is a 2 times reduction in the number of standard cells because single-height type A-B condition is the same as single-height type B-A condition.

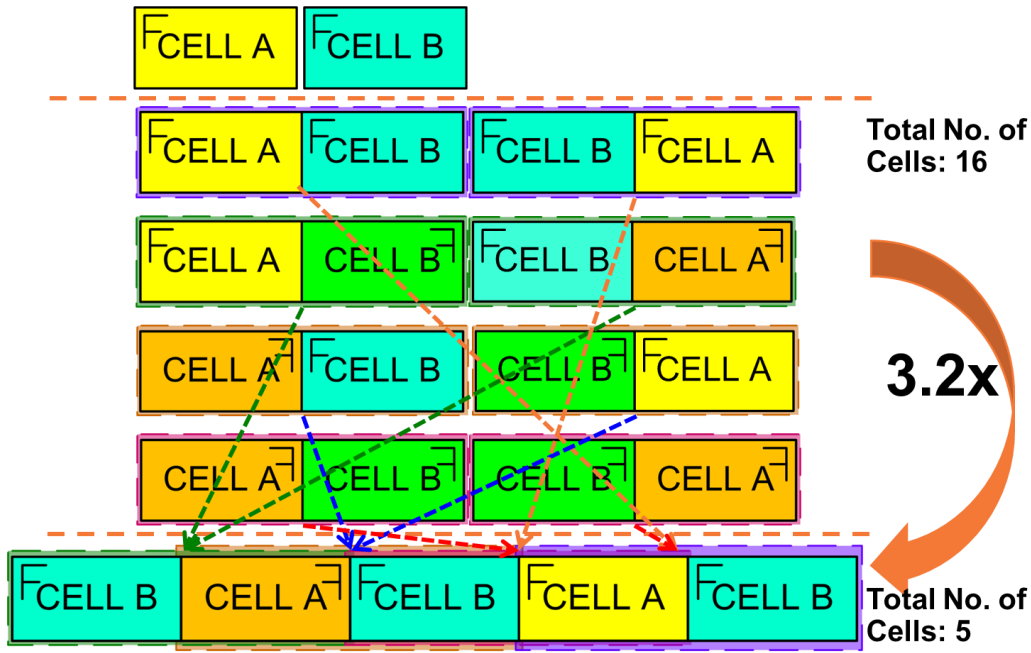

Figure 9. Proposed single-height type A-B cell abutment placement

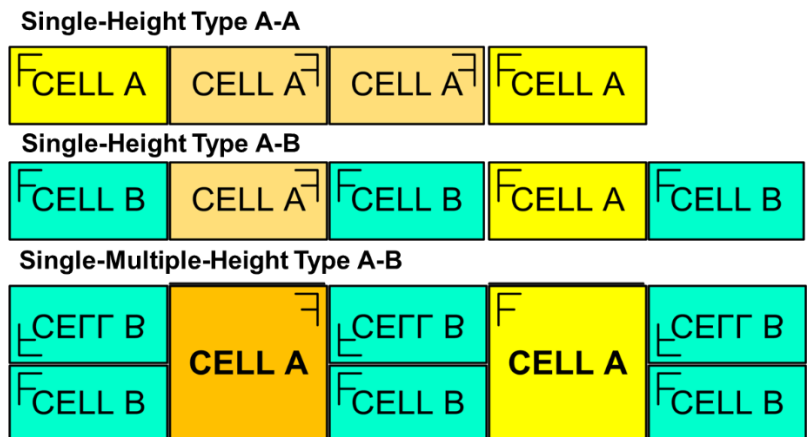

Figure 10. The proposed area-efficient cells abutment placement

In addition to single-height cells abutment, we have considered multiple-height cells in the standard cell library. Multiple-height cells such as level shifter cell are designed with an integer multiple of the single-height standard cells. Therefore, we have to consider two additional placement conditions, which are single-and-multiple-height placement and multiple-multiple-height placement. The single-height Type A-A can be applicable to multiple-height Type A-A placement. For single-height and multiple-height cell placement, we extended the proposed abutment placement from single-height Type A-B to single-multiple-height Type A-B placement, as shown in Figure 10. Without the loss of generality, this work did not consider multiple-height-multiple-height Type A-B, however, the methodologies can be extended from the single-multiple-height Type A-B abutment placement.

The breakdown of different scenarios allows us to further optimize the number of cells required for each abutment condition. To illustrate the benefits of our proposed method, let us consider a standard cell library with N number of single-height cells. Our method reduces the total number of cells from $8*N + 8*(N-1)*N$ to $4*N + 2.5*(N-1)*N$, whereas the Synopsys 'testcell' method only reduces to $6*N + 6*(N-1)*N$ *cells*. In one of our 14-nm technology IP library, the number of single-height standard cells is approximately 1000 cells. The conventional method requires a total of 8,000,000 cells to achieve hundred-percent coverage. The total number of cells used in the Synopsys 'testcell' method and our method is 6,000,000 and 2,505,500 cells, respectively. Therefore, our method provides close to 2.4 times reduction compared to the Synopsys 'testcell' method. Table 1 summarizes the total number of cells for each method.

Table 1.  The number of cells required for each method

| Number of Cells | Conventional Method | Synopsys 'testcell' | Our method |
|---|---|---|---|
| N | $8*N + 8*(N-1)*N$ | $6*N + 6*(N-1)*N$ | $4*N + 2.5*(N-1)*N$ |
| E.g 1000 | 8,000,000 | 6,000,000 | 2,505,500 |
| Reduction | 1x | 1.33x | 3.2x |

# 5  Design Implementation

The implementation of multiple lithography-compliant standard cell library validation flow is based on a customized Tcl procedure script, which leverages on the Synopsys IC Compiler commands [8]. This is a required tool to help layout engineers or IP engineers in checking and fixing potential lithography weak-points at the boundary of the standard cell abutment, especially during the library development phase.

In this work, we aim to simplify the use-model for the validation flow. As such, this Tcl procedure only requires user to provide the list of milkyway libraries as input. The simple-to-use approach reduces the barriers to adopting of new tool, especially our target users who might not be well-versed with programming and digital design flow.

The associated IC Compiler commands to implement the flow is summarized in Figure 11 and the customized Tcl procedure can be broken down into the following phases.
   A. Standard cell profiling
   B. Reduction techniques for standard cell abutment placement
   C. Verilog and DEF files generation
   D. Design Rule Verification (Include color-decomposition verification) and DRC+ Verification.

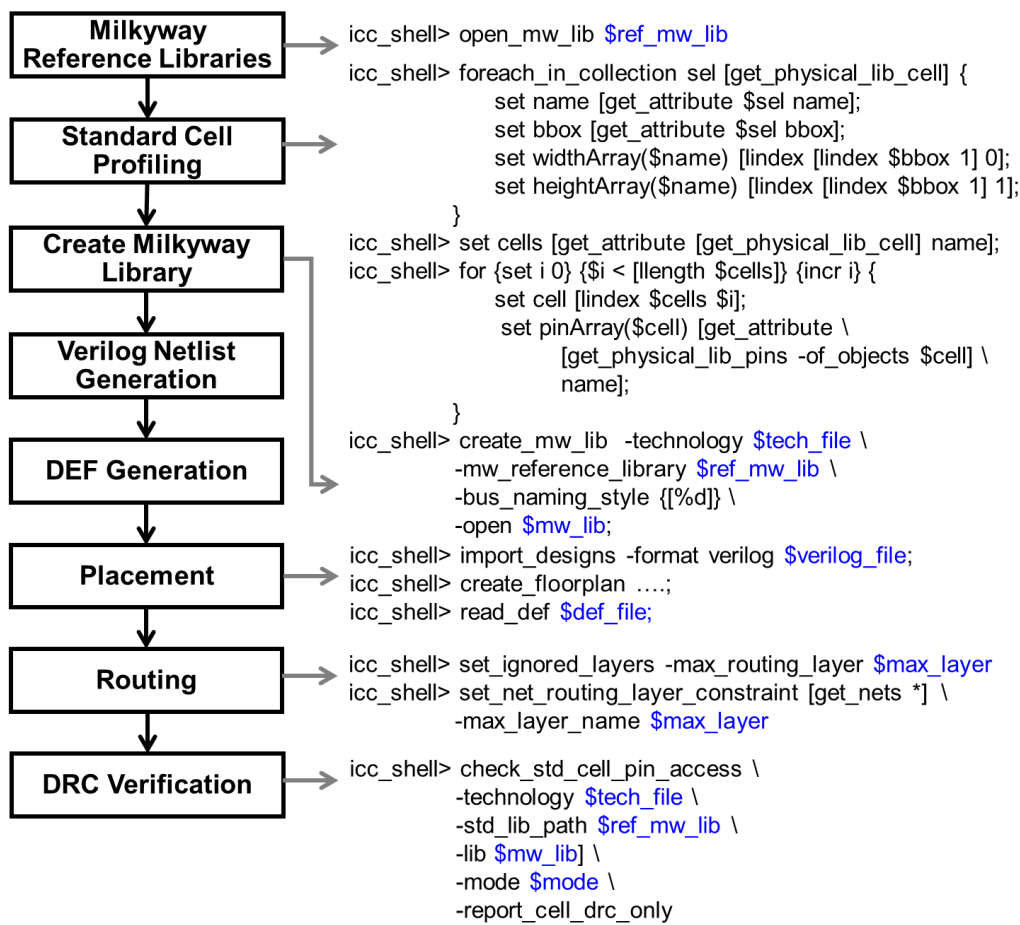

Figure 11. Workflow of our utility tool with the associated ICC commands.

*Standard Cell Library Evaluation with Multiple-lithography-compliant verification and Improved Synopsys Pin Access Checking Utility*

To enumerate all the standard cells abutment conditions, we need the basic standard cell information from the list of libraries. This can be done with the following algorithm:

```
array set widthArray {}; # Initial the array
array set heightArray {}; # Initial the array
array set pinArray {}; # Initial the array

foreach lib $mw_libs {
    open_mw_lib $lib; # Open the milkyway database;
    foreach_in_collection sel [get_physical_lib_cell] {
        # The standard cell physical information is stored in the Tcl array.
        set name [get_attribute $sel name]; # Get the cell name;
        set bbox [get_attribute $sel bbox]; # Get the cell boundary information;
        set widthArray($lib,$name) [lindex [lindex $bbox 1] 0]; # Get the cell width
        set heightArray($lib,$name) [lindex [lindex $bbox 1] 1]; # Get the cell height
    }
    # The standard cell pin information is stored in the Tcl array.
    set cells [get_attribute [get_physical_lib_cell] name];
    for {set i 0} {$i < [llength $cells]} {incr i} {
        set cell [lindex $cells $i];
        set pinArray($lib,$cell) [get_attributes [get_physical_lib_pins –of_objects $cell] name];
    }
}
```

The algorithm translates the standard cell physical information into Tcl arrays, which will be used for the Verilog netlist and design exchange format (DEF) file generations.

## 5.1   Verilog netlist and DEF file generation.

To implement the proposed area-efficient 'testcell' abutment placements in the Synopsys IC Compiler, we have adopted the netlist and DEF placement flow.

The Verilog netlist describes the logic cell required the 'testcell' abutment and all the different types of abutment placement can be generalized in the illustration shown in the Figure 12.  For example, for the single-height type A-A placement, it requires four identical cells, which is described under the module scell_<typeA> Verilog netlist.

The DEF file describes the physical layout of the 'testcell' abutment. It requires the basic standard cell information such as width and height, which is stored in the Tcl arrays during the standard cell profiling phase. The orientation of each cell is derived based on the type of abutment placement. For example, as shown in Figure 13, for the single-height type A-A placement, the cell orientation is N – FN – FN – N.

The Tcl procedure generates the Verilog netlist and DEF files according to the different types of abutment placement. The top level Verilog module includes all the abutment permutations to ensure 100% coverage of standard-cell placement topologies.

**Single-Height Type A-A**
**Multiple-Height Type A-A**
module scell_<typeA> ();
    <typeA> U1 (.*);
    <typeA> U2 (.*);
    <typeA> U3 (.*);
    <typeA> U4 (.*);
endmodule

**Single-Height Type A-B**
module scell_<typeA>_<typeB> ();
    <typeB> U1 (.*);
    <typeA> U2 (.*);
    <typeB> U3 (.*);
    <typeA> U4 (.*);
    <typeB> U5 (.*);
endmodule

**Single-Multiple-Height Type A-B**
module mcell_<typeA>_<typeB> ();
    <typeB> U1 (.*);
    <typeB> U2 (.*);
    <typeA> U3 (.*);
    <typeB> U4 (.*);
    <typeB> U5 (.*);
    <typeA> U6 (.*);
    <typeB> U7 (.*);
    <typeB> U8 (.*);
endmodule

**Top Module**
module TOP ();
    scell_<typeA> U1 ();
    scell_<typeA>_<typeB> U1 ();
    mcell_<typeA>_<typeB> U2 ();
endmodule

Figure 12. Verilog netlist for different standard cell abutment.

Example of the generated DEF file

```
VERSION 5.6 ;
DESIGN TOP;
TECHNOLOGY ROUTE ;
UNITS DISTANCE MICRONS 1000 ;
COMPONENTS 4;
 - sinst_<typeA>VU1 <TYPEA> + PLACED ( 0 0 ) N ;
 - sinst_<typeA>VU2 <TYPEA> + PLACED ( 200 0 ) FN ;
 - sinst_<typeA>VU3 <TYPEA> + PLACED ( 400 0 ) FN ;
 - sinst_<typeA>VU4 <TYPEA> + PLACED ( 400 0 ) N ;
END COMPONENTS
DIEAREA ( 0 0 ) ( 600 0 ) ;
END DESIGN
```

Figure 13. DEF file to define the placement of the 'testcell' abutment.

The following IC Compiler commands detail the floorplan implementation of the abutment placement:

```
create_mw_lib \
    -technology $tech_file \
    -mw_reference_libraray $mw_libs \
    -bus_naming_style {[%d]} \
    -open $lib; # Create the milkyway library
import_design –format Verilog $verilog_file; # Read in the Verilog file
create_floorplan …; # Create the floorplan (Optional)
read_def $def_file; # Read in the DEF file
```

## 5.2 Pin connectivity assignment

The pin connectivity assignment in the Synopys 'testcell' is illustrated in Figure 5. The pins of the cell A and B are connected to the corresponding pins in the cell A and B, respectively. Since the cells in the Synopsys 'testcell' are placed in a single row with either R0 or MY orientations, the pins are aligned in the same routing tracking. Due to its simplicity in the Synopsys 'testcell' pin connection, the physical signal routes in all the four modes are very similar. This is evidently shown in Figure 14 the layers ME2 and ME3 routed nets in (a) single_cell_only mode are almost identical to (b) all_combo_in_one_cell mode, except for the net highlighted in 'red' box. In our opinion, Synopsys PAC utility verification modes (1) and (2) would be sufficient for the pin-accessibility assessment.

Figure 15 details the layout view of the Synopsys 'testcell' and the routed signal layers. We have observed that each signal only requires two to three via transitions to connect the pins together. Since the horizontal distance between the two pins are within two to three cell's width apart and the pins are aligned in the same routing track, this will be a trivial solution for the router tool. To increase the routing congestion and to exacerbate the pin accessibility issue, we propose to randomly connect the pins in our 'testcell'. This increases the routing distance and the signal routes are crisscrossed all over the 'testcell'. The routed 'testcell' design will be close resemblance to routings in the real design.

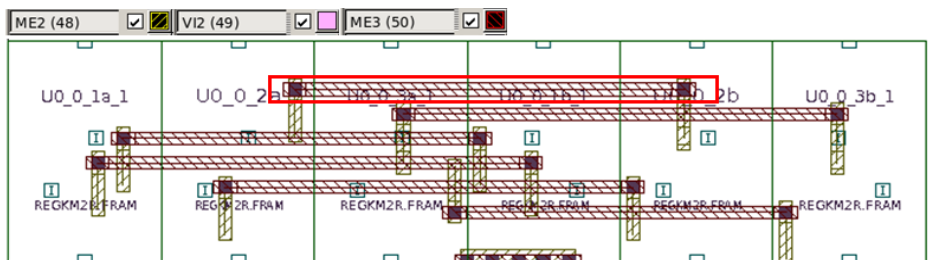

(a) ME2&ME3 routed nets with VE2 vias locations in single_cell_only mode

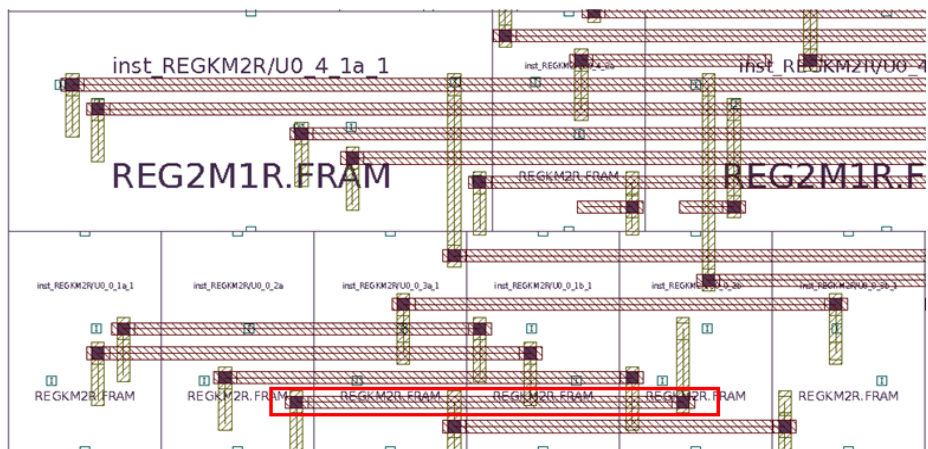

(b) ME2&ME3 routed nets with VE2 vias locations in all_combo_in_one_cell_only

Figure 14. An example of the Synopsys 'testcell' in the Synopsys layout environment (a) single_cell_only mode, (b) all_combo_in_one_cell.

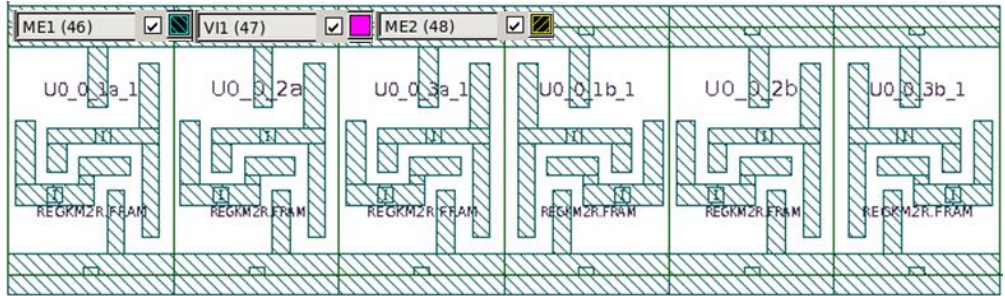

(a) Synopsys 'testcell' Placement

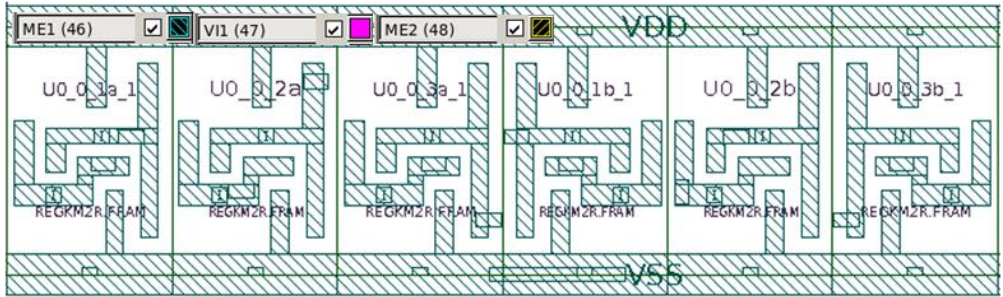

(b) ME1 routed nets

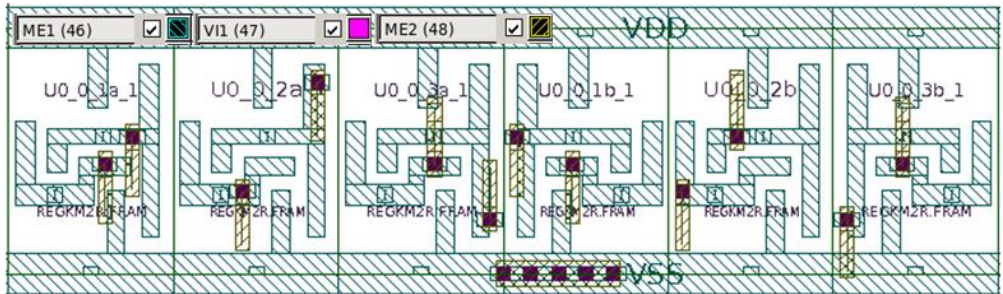

(c) ME1&ME2 routed nets with VE1 vias locations

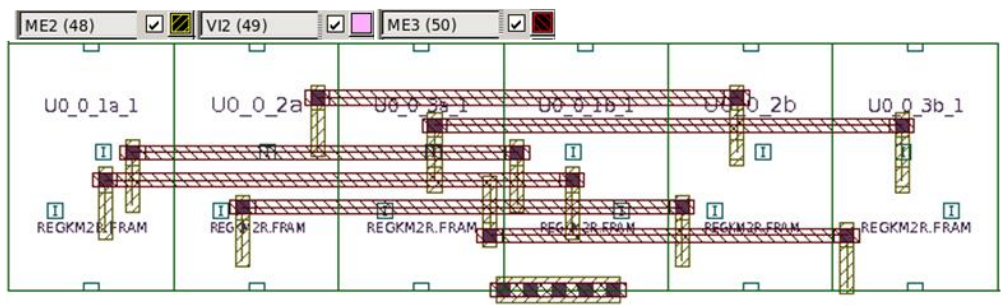

(d) ME2&ME3 routed nets with VE2 vias locations

Figure 15. An example of the Synopsys 'testcell' in the Synopsys layout environment (a) cell view of the Synopsys 'testcell', (b) ME1 layer signal routes, (c) ME2 and VE1 layers are generated on the pins during the routing phase. (d) ME3 and VE2 layers are generated to connect the pins together.

## 5.3  Routing Constraint

The technology file details the physical design rule constraints for the metal routing lines. To increase the routing congestion and to exacerbate the pin accessibility issue, Synopsys PAC utility provides the option 'route_option_file' that specifies the route options to be honored during the routing phase [9]. However, there is no clear guideline for the list of routing constraints needed.

In this work, we have experimented with various ICC routing constraints commands to create random blockage and routing nets to increase the routing congestion. This is list of commands can be used in the Synopsys PAC utility. The following ICC commands detail the routing constraints:

Step 1: Derive the routing layers information.
```
set layers [get_attribute [get_layers -filter is_routing_layer==true] full_name];
set horizontal_layer [get_attribute [get_layers -filter preferred_direction==horizontal] full_name];
set vertical_layer [get_attribute [get_layers -filter preferred_direction==vertical] full_name];
set via_layers [get_attribute [get_layers –filter is_routing_layer==true –filter layer_type==via] \
full_name];
foreach layer $layers { if {[lsearch $via_layers $layer] == -1} { append route_layers "$layer "; }; }
```

Step 2: Connects the power rails together.
```
derive_pg_connection -power_net VDD -power_pin VDD -ground_net VSS -ground_pin VSS;
preroute_standard_cells -fill_empty_rows \
    -advanced_via_rules \
    -nets "VDD VSS" \
    -route_type {P/G Std. Cell Pin Conn} \
    -extend_to_boundaries_and_generate_pins;
```

Step 3: Constraints the router to 2 routing layers.
```
set min_layer [lindex $layers 2]; # minimum routing layer
set max_layer [lindex $layers 4]; # maximum routing layer
set_preroute_drc_strategy -min_layer $min_layer -max_layer $max_layer;
set_ignored_layers -min_routing_layer $min_layer;
set_ignored_layers -max_routing_layer $max_layer;
set_net_routing_layer_constraint [get_nets *]  -min_layer_name $min_layer -max_layer_name $max_layer;
set constraints "";
for {set i 0} {$i <= [llength $route_layers]} {incr i } {
    if $i {
        append constraints "{[lindex $layers $i] false} "
    } else {
        append constraints "{[lindex $layers $i] true} "
    };
};
set_route_zrt_common_options  -freeze_layer_by_layer_name $constraints;
```

Step 4: Create random power straps (acts as random routing blockage).
```
set vertical_layer [get_attribute [get_layers -filter preferred_direction==vertical] full_name];
set ymax [lindex [lindex [get_attribute [get_die_area] bbox] 1] 1]; # size of the 'testcell'
set xmax [lindex [lindex [get_attribute [get_die_area] bbox] 1] 0]; # size of the 'testcell'
```

*Standard Cell Library Evaluation with Multiple-lithography-compliant verification and Improved Synopsys Pin Access Checking Utility*

```
foreach layer $route_layers {
    if {[lsearch $horizontal_layer $layer] >= 0} {
        create_power_straps -direction horizontal \
            -start_at 0 -stop $ymax \
            -layer $layer -nets "VDD VSS" \
            -width [expr int(rand()*100)/500.000] \
            -step  [expr int(rand()*100)/50.000] \
            -configure step_and_stop \
            -extend_low_ends force_to_boundary_and_generate_pins \
            -extend_high_ends force_to_boundary_and_generate_pins \
            -keep_floating_wire_pieces
    } else {
        create_power_straps -direction vertical \
            -start_at 0 -stop $xmax \
            -layer $layer -nets "VDD VSS" \
            -width [expr int(rand()*100)/500.000] \
            -step  [expr int(rand()*100)/50.000] \
            -configure step_and_stop \
            -extend_low_ends force_to_boundary_and_generate_pins \
            -extend_high_ends force_to_boundary_and_generate_pins \
            -keep_floating_wire_pieces
    }
}
```

Step 5: Constraints the router to fix open and short errors only.

```
set_route_zrt_detail_options -report_ignore_drc [list "Diff net spacing" "End of line spacing" \
    "Diff net var rule spacing" "Same net spacing" ... ]; # List of drc constraints to be ignored
    during reporting
```

To specify the router to fix the open and short locations, we use the following ICC commands:

```
route_zrt_detail -incremental true -initial_drc_from_input true
```

In this work, we have used the option 'report_cell_drc_only' in the Synopsys PAC utility to summarize the DRC error types. A simple experiment is conducted on one problematic standard cell. As shown in Figure 16, the Synopsys PAC's DRC report only indicates that there is one DRC error in our proposed 'testcell'. This demonstrates that our proposed methods have improve the odds of detecting the pin accessibility issues in the standard cell. As shown in Figure 17, in our proposed 'testcell', cell_REGKM2R_yf, there is a ME1 spacing violation at the standard cell boundary due to the insertion of the metal 2 and via 1 at the pin location. The ME1 spacing violation is highlighted with the 'red' box.

It is noted that the accuracy of the DRC report is dependent on the layout content in the milkyway database's FRAM view. For example, if the FRAM view only contains the pin information without M1 metal layer, the extra M1 metal layers added to fulfil the enclosure condition might violate the M1 metal layer spacing violation at the cell boundary. Therefore, it is highly encourage for engineer to use the complete layout in the FRAM view for the verification.

*Standard Cell Library Evaluation with Multiple-lithography-compliant verification and Improved Synopsys Pin Access Checking Utility*

```
=========================================================================
SCRIPT-Info: Printing DRC Summary ....
=========================================================================
########## 4 cells without DRC errors ##########
-------------------------------------------------------------------------
Cell                  DRC count  Master Cells with DRC              DRC Types
-------------------------------------------------------------------------
auto__icGRt           0
auto__icTrkAsgn       0
auto__icDRt           0
cell_REGKM2R          0
########## 1 cells with DRC errors ##########
cell_REGKM2R_yf       1                                            {Same net via-cut spacing}
```

Figure 16. Synopsys PAC utility's DRC report, reporting DRC errors in the proposed 'testcell' compared to the Synopsys 'testcell'.

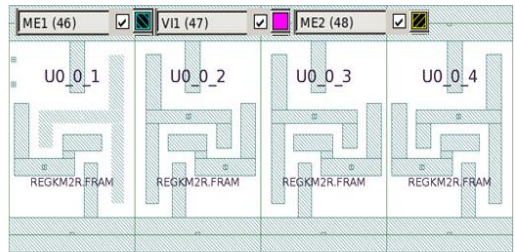

(a) Proposed 'testcell' Placement

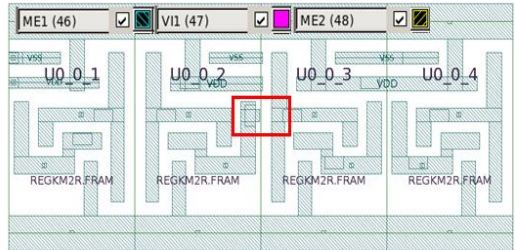

(b) ME1 routed nets

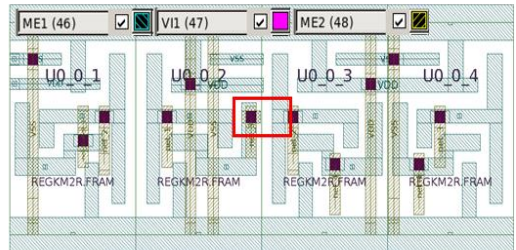

(c) ME1&ME2 routed nets with VE1 vias locations

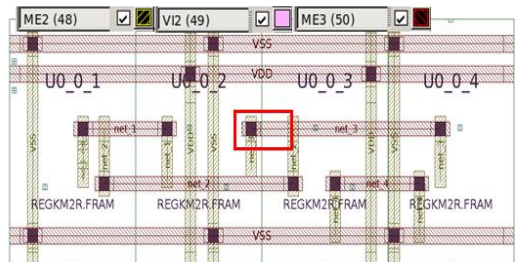

(d) ME2&ME3 routed nets with VE2 vias locations

Figure 17. An example of our 'testcell' in the Synopsys layout environment. The ME1 spacing violation is highlighted with the 'red' box. (a) Cell view of our 'testcell', (b) ME1 layer signal routes, (c) ME2 and Vi1 layers are generated on the pins during the routing phase. (d) ME3 and Vi2 layers are generated to connect the pins together.

## 5.4   Design Rule Check (DRC) and DRC+ Verifications.

After the placement and routing of the standard cells in the IC Compiler LayoutWindow, our utility provides the following physical verifications and summarizes the errors at the abutted 'testcell' boundaries.
- Design Rule Check (DRC)
- GLOBALFOUNDRIES Pattern matching (DRC+)

The IC Validator In-Design feature in IC Compiler provides the ability to use the IC Validator tool to perform the DRC and DRC+ physical verifications [9].

The following IC Compiler commands detail the IC Validator setup and perform physical verifications in IC Compiler:

```
# DRC Physical Verification
if {[file exists $drc_runset]} {
    set_physical_signoff_options –exec_cmd {icv} –drc_runset {$drc_runset}; # DRC Setup
    report_physical_signoff_options; # Report the DRC Setup
    signoff_drc –check_all_layers –run_dir {$result_dir} ; # DRC Setup
    process_drc_report –mode drc –run_dir $result_dir;
}

# DRC+ Physical Verification
if {[file exists $drcplus_runset]} {
    set_physical_signoff_options –exec_cmd {icv} –drc_runset {$drc_runset}; # DRC Setup
    report_physical_signoff_options; # Report the DRC+ Setup
    signoff_drc –check_all_layers –run_dir {$result_dir} ; # DRC+ Setup
    process_drc_report –mode drcplus –run_dir $result_dir;
}
```

*Standard Cell Library Evaluation with Multiple-lithography-compliant verification and Improved Synopsys Pin Access Checking Utility*

## 6    Result & Discussion

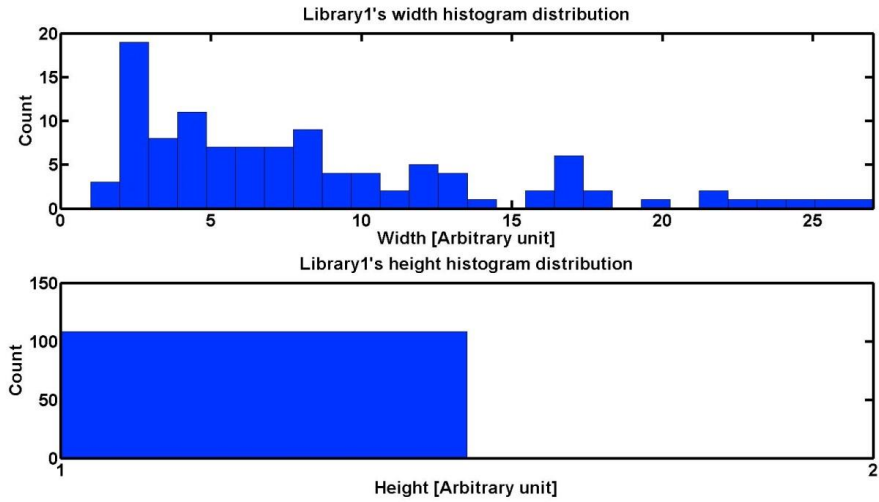

Figure 18. Normalized histogram distribution for the 9-track standard cell library1

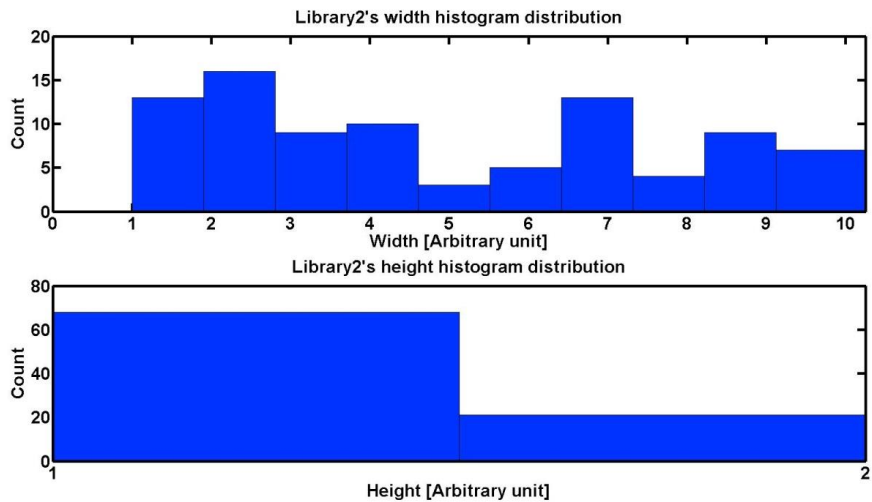

Figure 19. Normalized histogram distribution for the 9-track mixed-threshold standard cell library2

The experiment was carried out on five sets of 14-nm standard cell IP libraries. The library information is listed in the Table 2. The first four set of libraries belong to 9-track height design, which is optimized for mainstream design while the remaining library belongs to 10.5-track height design, which is optimized for speed critical design. 'library1' and 'library5' belongs to the standard threshold standard cell IP library. 'library2', 'library3' and 'library4' belongs to different engineering versions of the mixed-threshold standard cell IP library. As shown in Figure 18, 'library1' contains about eighty percent of the standard cells fall within ten times of the minimum standard cell width. As shown in Figure 19, the mixed-threshold standard cell IP library, 'library2' contain about thirty percent of multiple-height standard cells.

In our experiment, the metric used to compare the two utilities are the run-time, the size of milkway database and the number of problematic standard cells. The run-time refers to the total compute time taken for the verification to complete on our Linux workstation with Intel 2.7-GHz 8 Core Duo CPU and 128-GB of memory while restricting the router to a four thread for consistency in comparison. We constraints the routing to two metal layers (M2 and M3) and the routing directions of M2 and M3 are horizontal and vertical, respectively.

## 6.1 Lithography verification

For our first experiment, we have evaluated two possible scenarios where designers can either use the pre-defined coloured standard cell layout (DPT Option I) or to perform DPT using the Synopsys ICV DRC kit (DPT Option II).

We only observed lithography violations using the DPT Option I because DPT option II provides an additional opportunity to correct any coloring conflict after layout is being generated from the ICC. A common coloring violation location at the boundary of standard cells is illustrated in Figure 20, where metal lines bridging can occur due to two M1_E1 and two M1_E2 metal lines running in parallel.

Designers have to be aware that standard cells' spice models are highly dependent on the layout coloring. Even though DPT Option II is able to avoid any coloring violation, it might affect the circuit performance. Therefore, it is highly advisable for IP designers to fix the entire DRC+ violations before releasing the IP to their customers. The IP designers can use our utility to identify all the possible abutment conditions.

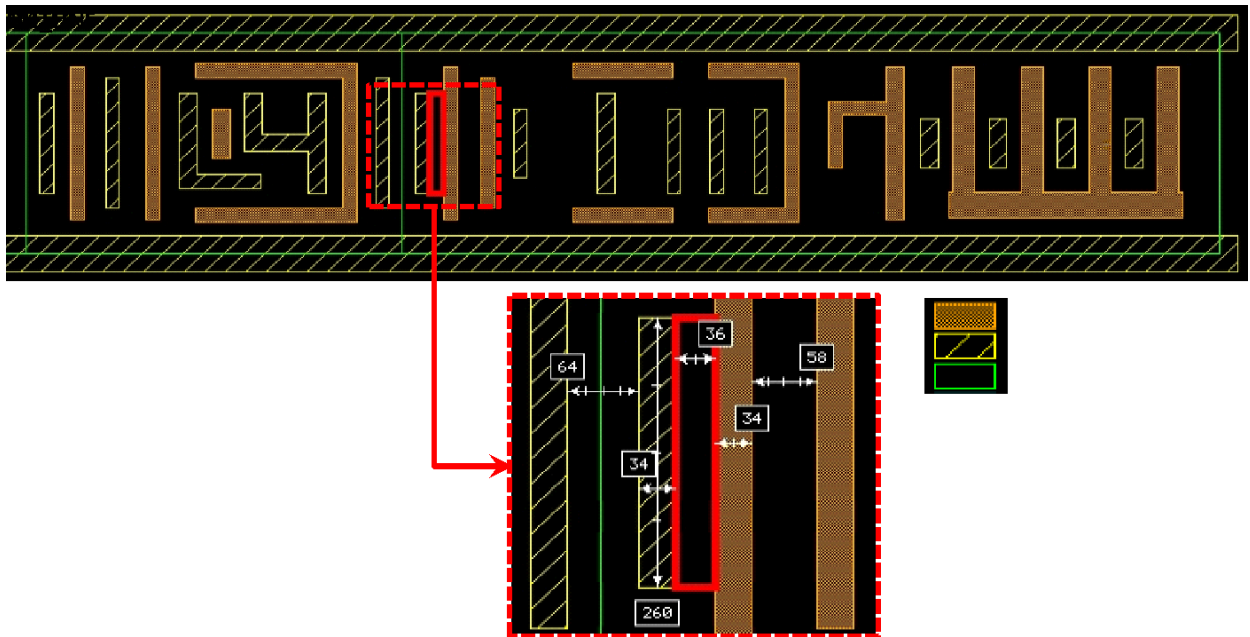

Figure 20. DRC+ hotspot found at boundary for standard cells that are fixed coloring before abutment placement (DPT Option I).

## 6.2 Pin accessibility assessment

For our second experiment, the benchmark information for evaluating the quality of standard cell library through pin accessibility check is summarized in Table 2. From the result, our proposed utility is able to identify more problematic standard cells. As expected, there will be more problematic cells in the 9-track libraries compared to the 10.5-track due to the decreased opportunity in the routing space and pin placement in the former libraries. In general, we have noticed that most of the DRC violations can be categorized under spacing violations. Examples of the DRC violations occurred at the pin locations are illustrated in Figure 21 (Local double pattern cycle violation) and Figure 22 (Diff net spacing).

Table 2. Comparison result between our proposed method and present recommended method.

| Library Name | Recommended methodology presented in [7] | | | Our proposed method | | | |
|---|---|---|---|---|---|---|---|
| | No. of cells with remaining DRC Route issues | Run-time [hours] | Database [Mb] | No. of cells with remaining DRC Route issues | Run-time [hours] | Database with incremental result [Mb] | Database without incremental result [Mb] |
| library1 | 3 | 5.6 | 218 | 14 | 106 | 2,700 | 155 |
| library2 | 6 | 5.2 | 147 | 28 | 215.5 | 5,400 | 329 |
| library3 | 6 | 5.3 | 167 | 16 | 195.5 | 3,800 | 236 |
| library4 | 6 | 5.5 | 182 | 11 | 192.5 | 3,600 | 234 |
| Library5 | 0 | 4.93 | 222 | 21 | 80 | 503 | 190 |

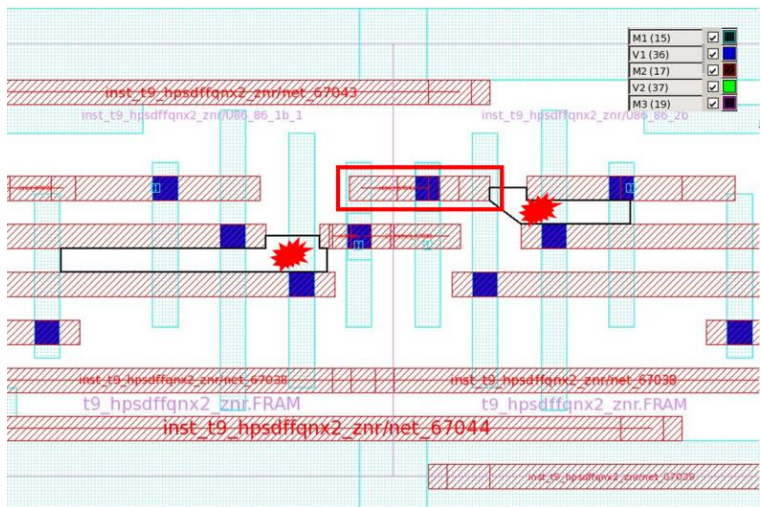

Figure 21. An example illustrating the odd pattern cycle errors near the pin locations due to the insertion of the metal 2 (M2) and via (V1) layers for routing.

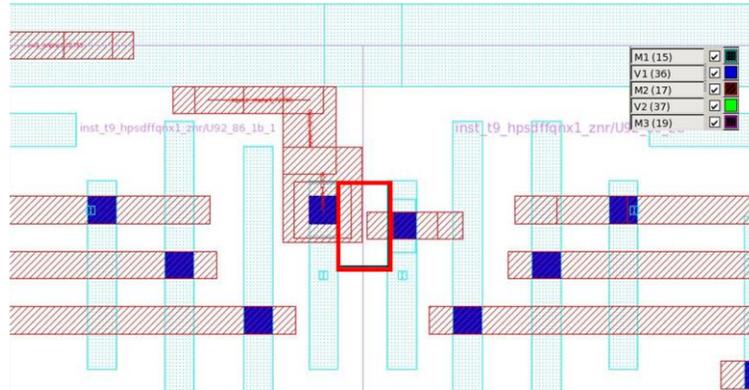

Figure 22. An example illustrating metal 2 (M2) spacing errors due to the insertion of the metal 2 (M2) and via (V1) layers for routing.

# 7   Future Work

Beside the proposed improvements discussed in Section IV, we have identified opportunities to enhance the Synopsys PAC methodology in our future work. One example is to modify the milkyway technology file. When we increase 'minEnclosedWidth' to the recommended design margin for better design robustness and manufacturability, we can identify and correct problematic standard cells based the pin locations with the spacing DRC violations. Similarly, when we increase the 'minSpacing' and 'minWidth' to the recommended design margin, it creates routing congestion and enhances the probability of detecting pin accessibility issues.

# 8   Conclusions

In this work, we have demonstrated our easy-to-use utility for the library physical layout designers to validate the design- and lithography-compliances of their standard cell library through the Synopsys IC Compiler. To achieve 100% standard cell abutment coverage with minimum verification runtime, we have proposed an area-efficient abutment placement, which achieves a 3.2 times reduction compared to the conventional single-height side-side standard cell abutments. The technique significantly reduces the runtime, memory and minimizes redundant violation errors. We also proposed several enhancements to the Synopsys PAC methodology and improved the probability of detecting pin accessibility issues. The significance of the presented methods helps library physical layout designers to perform extensive physical verification and identify opportunities to improve design manufacturability before silicon validation phase. As a result, we are expecting a better quality of result for the standard cell design, enabling our customers to achieve best-in-class PPA for their design.

# 9   Bibliography

[1] Kawa, C. C. Chiang, et. al., "Design for Manufacturability and Yield for Nano-Scale CMOS", Series on Integrated Circuits and Systems Springer, 2007.

[2] B. P. Wong, et. al., "Nano-CMOS Design for Manufacturability: Robust Circuit and Physical Design for Sub-65Nm Technology Nodes", New York, NY, USA: Wiley-Interscience, 2008.

[3] M. K. Hsu, et. al., "Design and manufacturing process co-optimization in nano-technology (Designer Track Paper)", in *IEEE/ACM International Conference on Computer-Aided Design (ICCAD)*, San Jose, CA, 2014.

[4] T. Taghavi, et. al., "New placement prediction and mitigation techniques for local routing", in *In Proceedings of IEEE/ACM International Conference on Computer-Aided Design*, 2010.

[5] Marek-Sadowska, et. al., "Can pin access limit the footprint scaling?", in *In Proceedings of ACM/IEEE Design Automation Conference*, 2012.

[6] Synopsys, "Application Note for Physical Library Analysis for Optimal 28-nm and Below Process Node Routing (Version 5.0)", Solvnet Article, 2014.

[7] Olivier Aupoix, et. al., "Optimizing standard cell pin accessibility in 14nmFDSOI with Synopsys pin access checker", *Synopsys Users Group (SNUG) France*, 2014.

[8] Synopsys, IC Compiler Implementation User Guide Version, 2016.

[9] Synopsys, IC Compiler Advanced Geometries User Guide, 2016.

*Standard Cell Library Evaluation with Multiple-lithography-compliant verification and Improved Synopsys Pin Access Checking Utility*